\documentclass[pre,aps,superscriptaddress,a4paper,twocolumn]{revtex4}

\usepackage[utf8x]{inputenc}
\usepackage{graphicx, subfigure}
\usepackage{epstopdf}
\usepackage{amsmath}
\usepackage{amsfonts}
\usepackage{color}
\usepackage{tabularx}
\usepackage{textcomp}

\begin{document}
\title{Orientational ordering of lamellar structures on closed surfaces.}
\author{J. P\c ekalski}
\author{A. Ciach}
\affiliation{Institute of Physical Chemistry, Polish Academy of Sciences, Kasprzaka 44/52, 01-224 Warszawa, Poland}
%\date{\today}
\begin{abstract}
Self-assembly of particles with short-range attraction and long-range repulsion (SALR) interactions on 
a flat and on a spherical surface is compared. Molecular dynamics (MD) simulations are performed for 
the two systems having the same area and the density optimal for formation of stripes of particles.
Structural characteristics, e.g. a cluster size distribution, a number of defects and an orientational 
order parameter (OP), as well as the specific heat, are obtained for a range of temperature.  In both cases,
the cluster size distribution becomes bimodal and elongated clusters appear at the temperature
corresponding to the maximum of the specific heat. When the temperature decreases, orientational 
ordering of the stripes takes place, and the number of particles per cluster or stripe increases
in both cases. However, only on the flat surface the specific heat has another maximum at the temperature
corresponding to a rapid change of the OP. On the sphere, 
the crossover between the isotropic and anisotropic structures occurs in a much broader temperature interval,
the orientational order is weaker,
and occurs at significantly lower temperature. At low temperature the stripes on the sphere form spirals, 
and the defects resemble defects in the nematic phase of rods adsorbed at a sphere. 
\end{abstract}
\maketitle

\section{Introduction}

 Competing interactions of various origin can lead to pattern formation at different length
 scales~\cite{seul:95:0,gelbart:99:0,stradner:04:0,sedgwick:04:0,campbell:05:0,rauh:16:0,zhuang:16:0,zhuang:16:1,zhuang:16:2,edelmann:16:0,pini:17:0}.
 In particular, particles adsorbed at flat interfaces can form  stripes or spherical clusters or bubbles,
 when the hard-core 
 repulsion is followed by attraction at larger distances, and a repulsion is again present at still larger 
 particle separations (SALR potential) \cite{imperio:04:0,archer:08:0,schwanzer:10:0,roth:11:0,chacko:15:0,almarza:14:0,pekalski:14:0}.
 For increasing density of the particles, transitions between disordered
 fluid, hexagonal cluster crystal, stripes, hexagonal crystal of bubbles and dense disordered phase 
 occur at sufficiently low temperature $T$~\cite{almarza:14:0,pekalski:14:0}.  
 
 For the density optimal for the stripe formation, an isotropic labyrinth of stripes occurs 
 at relatively high  $T$. 
 This isotropic inhomogeneous phase is transformed to an anisotropic phase of stripes 
 with preferred orientation (molten lamella, ML) when  $T$ is decreased~\cite{almarza:14:0}.
 In a SALR system on a triangular lattice, further decrease of temperature leads to the transition to the lamellar
 phase (L) with a translational order, as found (for a finite system) by Monte Carlo simulations in 
 Ref.\cite{almarza:14:0}.
 
  A transition between the isotropic and anisotropic phases of stripes was observed also in thin magnetic 
  films~\cite{saratz:10:0}, and found in Ref.~\cite{barci:13:0}
 within the Landau-Brazovskii (LB) theory~\cite{brazovskii:75:0}. 
 The phenomenological LB theory is
 applicable to any system with competing interactions leading to 
 the order parameter (OP) that oscillates in space. The SALR model is a 
 particular member of the LB universality class~\cite{ciach:13:0}. One may expect that some of the results
 concerning the structure of stripes obtained for the SALR model
 can be valid in other stripe-forming systems as well.
 
  The L and ML phases are analogous to the smectic and nematic liquid crystals respectively, 
 because in the L and smectic phases both the translational and the orientational order are present, 
 and the ML and nematic
 phases are ordered only orientationally. The translational order in those phases is
 destroyed by topological defects like dislocations and disclinations.
 
 The defects were intensively studied for stripe forming systems mainly in the case of block-copolymers 
 thin films~\cite{hammond:05:0,horvat:08:0,mishra:12:0}. In the case of the SALR system, 
 the defects can be suppressed 
 when the two-dimensional system is confined between two parallel boundaries (lines),
 i.e. decreasing the wall separation has an effect analogous to decreasing
 the temperature~\cite{almarza:16:0}. The geometry, permeability and elasticity of the confinement, 
 however, play an important role for the ordering 
 of inhomogenous SALR structures~\cite{imperio:07:0,roth:11:0,pekalski:15:0,pekalski:15:1}.
 
 In this work we focus on the SALR particles adsorbed at a surface of a vesicle or at a spherical droplet.
 Particles adsorbed at the surface of a droplet, i.e. colloidosomes, have been 
 studied mainly in the case of close 
 packing~\cite{geerts:10:0,rossier:09:0,yang:17:0,guan:16:0}. 
 Recently, such systems attract attention also in the case of the sphere only partially
 covered by the particles, because of the possibility of spontaneous pattern formation.
 %in the case of competing interactions. 
 In the case of DNA-coated colloids on functionalized oil droplets~\cite{joshi:16:0}, it was 
 shown that various patterns can be formed when one tunes the strength of the attraction between the colloids.
 The studies in Ref.\cite{joshi:16:0}, however, were limited to densities corresponding to formation of the bubbles, therefore the Authors
 did not find any lamellar or cluster morphology.
 
 There are no true thermodynamic phase transitions in finite systems, therefore we do not expect any 
 ordered phase on a surface of a sphere. However, the structure of the disordered phase 
 may be significantly different 
 for different densities and at different
 temperatures, due to a presence of a short-range order. We call the finite system ``ordered'', when
 the correlation length is larger than the size of the system, and a suitably defined OP
 is of order of unity. In this context, 'the ordered structure' should not be mistaken with 'the ordered phase'. 
 
 The global ordering on curved and closed surfaces is suppressed by defects that are inevitably 
 associated with the curvature. 
 The interesting  question of defect formation at curved surfaces attracted a lot of 
 attention recently  \cite{bowick:09:0,keber:14:0,perotti:16:0,allahyarov:17:0,azadi:16:0,irvine:10:0,gomez:15:0,zhang:14:0,guerra:18:0},
 but the studies focused mainly on closely-packed particles \cite{nijs:15:0,paquay:16:0}. 
 Studies of the pattern formation by the SALR particles on closed curved surfaces are 
  rare \cite{zarragoicoechea:09:0, amazon:13:0,goh:13:0,amazon:14:0,joshi:16:0}. 
  In the current study, we focus on colloidal particles with a diameter
  of a few hundreads of nanometers that are adsorbed on a spherical droplet with a diameter
 $\sim 10\mu m$, as in Ref. \cite{joshi:16:0}, i.e. the radius of the droplet is one order 
 of magnitude larger than the particle diameter.

 We cannot expect that at any temperature the stripes on a sphere can have the translational
 order such that the average 
 density is given by an oscillatory function $\rho(z)$ of a single scalar variable $z$, 
 as  in the lamellar phase on a flat surface. For this reason we do 
 not study the translational ordering. It is not clear, however, if
 the SALR particles on a surface of a sphere can self-assemble into stripes that
 have a preferred orientation, and how the orientational order can be quantified.
 %In this work we ask the question if  the orientational ordering of the SALR particles on a surface of a sphere can occur.
 As far as we know, this question has not been studied yet, and
 we address it in the present work.  In addition, we study if the orientational ordering is associated
 with the other structural characteristics such as the number of defects, the cluster size distribution, 
 the average size of the cluster and thermal properties (specific heat). Our studies are based on molecular
 dynamics simulation (MD).

 In the case of liquid crystals, the orientational OP is usually
 based on the eigenvalues of the ordering matrix \cite{AllenTildesleyBook}. This order parameter successfully
 describes nematic order in amphiphilic systems \cite{mccoll:72:0,lebwohl:72:0}. However, it is inconvenient
 in MD studies where structures,
even when ordered, flow and can change their orientation globally. To overcome this problem,
we propose a new orientational order parameter that
% by construction is independent of any coordinate system,  
is suitable for describing the order found by MD simulations, and is easily
applicable to systems with different topologies. We verify if the new OP gives results consistent with the 
standard OP based on the eigenvalue of the ordering matrix~\cite{AllenTildesleyBook} by calculating both parameters for a flat surface. 

In order to distinguish the effects of the finite size of the system from the effects of the curvature,
we consider a flat surface with periodic boundary conditions (torus topology, TBC) 
and a surface of a sphere (SBC) with the 
same area as that of the flat one. 
Likewise, to eliminate the effects of incompatibility between the number of particles
on the surface and the density 
optimal for the stripe formation, 
we fix the number of particles such that the density is close 
to the density optimal for the periodic lamellar structure, both on the flat and on the curved 
surface. On the curved surface
 the density 
optimal for the stripes is a bit smaller than on the flat one.
We choose the SALR potential leading to small clusters or stripes of thickness
$\sim 2\sigma$, where $\sigma$ is the particle diameter.

In Section \ref{sec:mm}, we define the model, introduce the orientational order parameter and describe 
the methods used for calculating
the size distribution of the aggregates and the average number of dislocation defects. 
In Section \ref{sec:torus}, we present the results for the flat surface. We show typical snapshots of the structure,
and temperature dependence of the following quantities:
the number of clusters or stripes,
the maximal number of particles in the aggregate, 
the two OP, 
the average number of defects,
the aggregate size distribution,
and the specific heat.
In Section \ref{sec:sphere}, we present the results for the same quantities for the 
particles adsorbed at the sphere.
In Section \ref{sec:summary} we discuss the results. 

\section{The model and the methods}
\label{sec:mm}
% Model
\subsection{Model}
\label{subsec:model}

The SALR potential is modeled as a sum of the Lennard-Jones and the Yukawa potentials:
\begin{equation}
V(r) = 4 \varepsilon \left[\left( \frac{\sigma}{r} \right) ^{2\alpha} - \left( \frac{\sigma}{r} \right) ^\alpha \right] + \frac{A}{r} e ^{-r/ \xi},
\label{Vc}
\end{equation}
where $\alpha = 6$, $A = 1.27$, $\xi = 2$, and $\varepsilon = 1.0$, $\sigma =1.0$ set the unit of energy and length, respectively.
{\color{black} In this case, the temperature has the units of $k_BT/\varepsilon$.} With that model parameters the area of the attraction well is around 3 times smaller than the area 
of the potential repulsive tail, i. e. $\int_1^{r_0} V(r) dr \approx 3 \int_{r_0}^{r_{cut}} V(r) dr$, 
where $r_0$ is the distance at which $V(r)$ crosses zero and $r_{cut}$ is the distance at which 
the potential was cut and shifted to zero for computational reasons. The used ratio matches the 
ratio of the repulsion to attraction strengths studied in the lattice model in Refs.~\cite{pekalski:15:0,pekalski:15:1,pekalski:14:0,almarza:14:0,pekalski:13:0}. The same form of the SALR potential was previously studied in Refs.  \cite{sciortino:04:0,sciortino:05:0,mani:14:0,toledano:09:0,santos:17:0}.

  \begin{figure}[h]
    \begin{center}
  \includegraphics[scale=1]{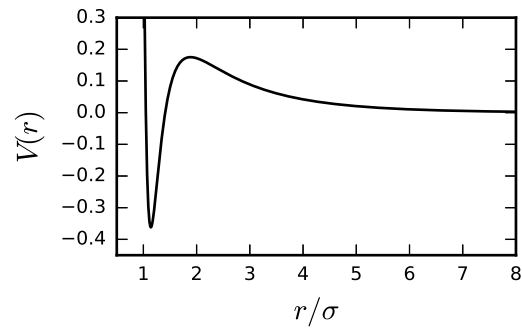}
      \caption{The potential has a local minimum for $r = 1.141\sigma$,
      then it crosses zero at $r_0 = 1.412 \sigma$ and it has a local maximum for $r = 1.886 \sigma$.
      The potential has been cut off at $r= 8\sigma$.
      }
      \label{ensemble_compar2a}
    \end{center}
  \end{figure}

\subsection{Methods}
\label{sec:methods}
% simulation method
In order to perform MD simulations, the HOOMD-blue~\cite{hoomd:1,hoomd:2}
package was used. We obtained the low temperature structures by linearly decreasing the temperature of the system.
Typically, the starting temperature was set to $k_BT = 0.2$ and the procedure finished at $k_BT = 0.03$, where
$k_B$ is the Boltzmann
constant. 
In all the runs at least $10^9$ MD steps were performed and the values of interest were sampled every $10^4$ step.
In order to obtain ordered structures, relatively small systems were considered, namely in the case of the 
toroidal periodic boundary conditions we used  $N=975$ particles, and a square simulation box with $L = 51.52\sigma$,
while in the case of the spherical boundary conditions, the number of particles was set to $N = 900$,
and the sphere radius was $13.78\sigma$. 
In the latter case the particles were constrained to move on the surface of the sphere.

% cluster formation analysis
The analysis of the formed aggregates was made based on the distance criterion with a cut-off
distance set to $r = 1.41\sigma$,
that is the distance where the pair potential crosses zero and becomes repulsive.
{\color{black} Using the width of the basin of attraction as a cut-off distance for identification of the aggregates  is a typical choice in SALR systems studies \cite{godfrin:13:0,godfrin:14:0,jadrich:15:0,bollinger:16:0,bomont:17:0,santos:17:0}}.
The aggregate size distribution (ASD) is presented in the manner commonly used in micellization studies,
where the probability of cluster occurrence is weighted by the cluster size, so that
\begin{equation}
\label{p(M)}
 p(M) = \frac{ P(M)M}{\sum_M P(M)M},
\end{equation}
 where $P(M)$ is the probability of finding an aggregate of size $M$.
 Since at high temperatures the particles form structureless aggregates,
 and then upon cooling larger elongated structures stabilize, we will refer to the former as 
 clusters and to the latter as stripes. At some temperature range, however, the clusters and  the stripes 
 coexist and the ASD is bimodal.  Thus, we can identify the maximal size of what we call cluster, $M_b$,
 with the minimum in the ASD that separates the two local maxima.  In the case of our model  $M_b = 22$.

\begin{figure}[t!]
    \begin{center}
	\centering
      \includegraphics[scale=1]{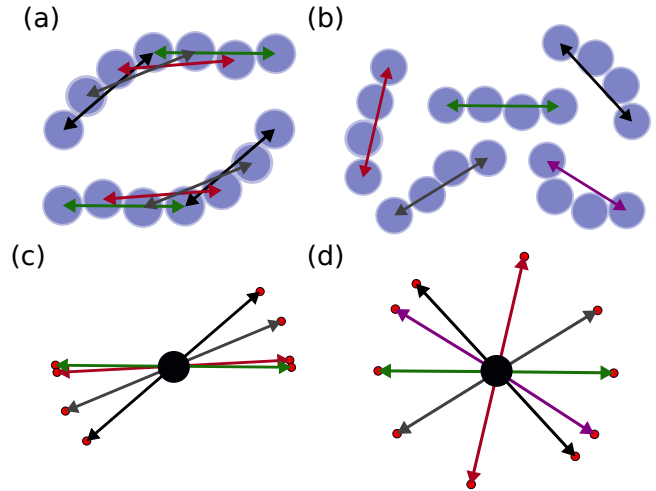}
      \caption{{\color{black} Illustration of the way the order parameter, $O_p$, is constructed. In panels (a-b) we show possible particle configurations at a flat surface and the vectors used for construction of $O_p$. After all the vectors are found, they are normalized and moved to one point (panels (c-d) respectively). The points indicated by the terminal points of the vectors (red circles) form an object which principal moments of inertia, $I_1$ and $I_2$ determine the value of the order parameter in the following way: $O_p = 1 - I_1/I_2$.}
      }
      \label{order_schem}
    \end{center}
 \end{figure}

\begin{figure}[t!]
    \begin{center}
	\centering
      \includegraphics[scale=1.5]{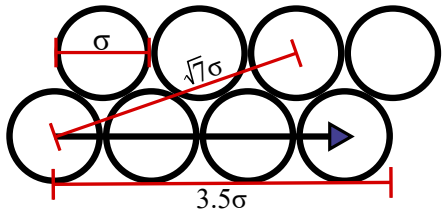}
      \caption{Schematic illustration of the construction of the orientational OP $O_p$. 
      The black vector representing the local orientation of the stripe, connects the centers of the particles
      located in the same row. The distance between the centers is assumed to be $\sqrt 7\sigma<r<3.5\sigma$. All such vectors are used
      in construction of the object, whose principal moments of inertia allow to distinguish isotropic and anisotropic structures.
      See the Section \ref{sec:methods} for more details. 
      }
      \label{vec}
    \end{center}
 \end{figure}

We determine if 
the stripes are orientationally ordered by considering local orientations of short segments within each of the lamellar stripes.
Namely, we find vectors that connect two particles within the same stripe, normalize them, drag all initial points of 
such vectors to one point, {\color{black} $O$}, and compute the moment of inertia of the object built by the vectors terminal points.
{\color{black} Because the considered particles are isotropic, the result should be independent of the {vectors\textquotesingle}  directions. Thus, we consider both directions of the vectors and as a results the built object is symmetric with respect to the point $O$ (see Fig. \ref{order_schem}). }
To make sure that each of the vectors describes the local orientation of the hosting stripe,
we take into account only pairs of particles belonging to the same row of the stripe. 
This is obtained by considering only the pairs of particles with the distance between their
centers larger than $\sqrt{7}\sigma$ and smaller than $3.5\sigma$ (see Fig. \ref{vec}). 

For a perfectly isotropic lamellar structure in 3D, the terminal points of the vectors should 
lay on a surface of a sphere, while for the a structure with a perfect orientational order (parallel stripes)
%which lay on spheres latitudes
 the terminal points would form a circle. Hence, we define the order parameter $O_p$ as
 {\color{black}one minus} the ratio 
between the largest and the smallest ones of the principal moments of inertia of this object,
{\color{black}  i. e. $O_p = 1 - I_1/I_2$}.

 For the case of TBC we compare $O_p$ with another orientational order parameter, 
 i.e. the eigenvalue, $\lambda$, of the ordering matrix\cite{AllenTildesleyBook}:
 \begin{equation}
 Q_{\alpha \beta} = {\color{black} \frac{1}{N_{\text{pair}}} } \sum_i {\color{black} 2} e_{i \alpha} e_{i \beta} - \delta_{\alpha \beta},
\label{order_mat}
  \end{equation}  
 where $\alpha$ and $\beta$ are the Cartesian coordinates in {\color{black} the two-dimensional space (2D)}, the
 index $i$ labels all {\color{black} $N_{\text{pair}}$ }considered normalized pairs of vectors and
 $\delta_{\alpha,\beta}$ is 
 the Kronecker delta function. {\color{black} It's important to note that, unlike in the 3d space, in 2d the ordering matrix has two eigenvalues $\pm \lambda$ which do not vanish when the fluid is isotropic and $N_{\text pair} < \infty$. \cite{frenkel:85:0}}

\begin{figure}[t!]
    \begin{center}
	\centering
      \includegraphics[scale=1.0]{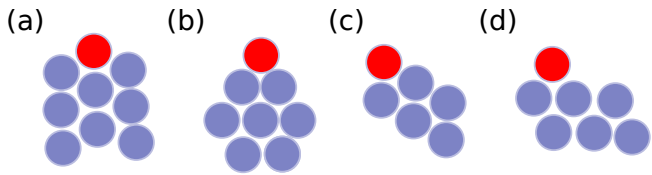}
      \caption{{\color{black} Representative configurations of particles considered as the dislocation defects. The particles marked by red color are the defects pointers located as described in the text.}
      }
      \label{defect_alg}
    \end{center}
 \end{figure}

% defect finding
We identify a dislocation defect in the stripe-forming system with breaking of the stripe, i.e. with
formation of two extra ends of the stripes. 
The stripes ends are found by analyzing the number of first 
and second neighbors of all the particles that form stripes. We take into account only particles with 2 or 3 
nearest neighbors (nn) and accept them as defect-pointers if the following conditions are fulfilled: a
particle has 3 nn and two of those nn have 3 nn whereas one of them has 6 nn {\color{black} (Fig. \ref{defect_alg}a)}, 
or a particle has 2 nn and both of them have 
4 nn  {\color{black}  (Fig. \ref{defect_alg}b)}, or one of them has 3 nn and the other one has 4 nn  {\color{black}  (Fig. \ref{defect_alg}c)}, or one has 3 nn and the other one 5 nn  {\color{black}  (Fig. \ref{defect_alg}d)}. 
The total number of dislocation defects, 
$d$, is defined as the total number of stripes ends divided by two, 
since breaking of one stripe results in two extra ends.

The visualization of the snapshots was made using The Open Visualization Tool (OVITO) \cite{ovito}.

\section{Results}
\label{sec:results}

The aim of our study is the comparison of the particle self-assembly into lamellar structures in systems with toroidal 
and spherical periodic boundary conditions. In both systems boundary conditions are periodic, however, the
toroidal boundary conditions (TBC) are used to mimic flat and infinite systems, whereas in the case of the spherical
boundary conditions (SBC) the surface is curved and closed. In what follows we present how the lamellar structures 
self-assemble in the two systems when the temperature is decreased.

\subsection{Toroidal boundary conditions}
\label{sec:torus}

  \begin{figure}[t!]
  \centering
      \includegraphics[scale=1.1]{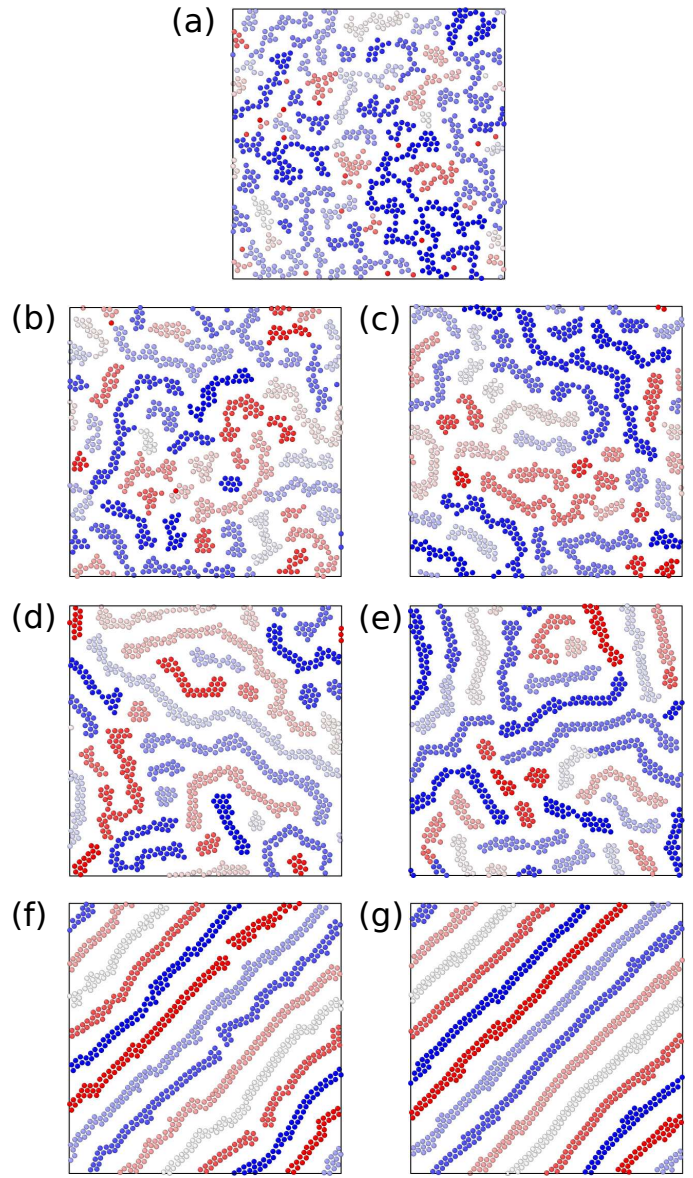}
            \caption{Representative configurations of the system with TBC and $N = 975$ at different temperatures. 
            (a) $k_BT = 0.25$, (b) $k_BT = 0.18$, (c) $k_BT = 0.14$, (d) $k_BT = 0.12$, (e) 
            $k_BT = 0.11$, (f) $k_BT = 0.10$, 
            (g) $k_BT = 0.05$. The number of clusters in the consecutive panels are: 80, 45, 38, 24, 30, 7, 7.
      }
      \label{torus:snaps}
  \end{figure}

\begin{figure}
    \begin{center}
	\centering
	\includegraphics[scale=1]{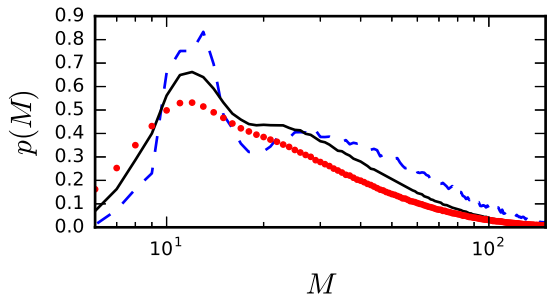}\\
            \caption{Mass weighted aggregate size distribution (Eq.(\ref{p(M)})) for the system with TBC, $N = 975$, shifted 
            vertically for better visibility. $k_BT$ = 0.20 (red dotted line), $k_BT$ = 0.165 (black solid line), 
            $k_BT$ = 0.12 (blue dashed line). Note the log scale on the horizontal axis.
      }
      \label{torus:csd}
    \end{center}
  \end{figure}

\begin{figure}
    \begin{center}
\includegraphics[scale=1]{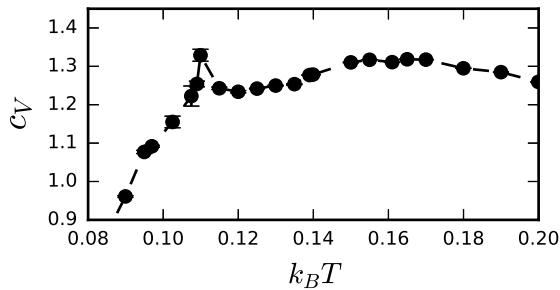}
            \caption{The Canonical heat capacity $c_V$ per particle as a function of temperature for the system with TBC. 
            The location of the high-$T$ maximum corresponds to the temperature at which the aggregate
            size distribution 
            becomes bimodal (Fig. \ref{torus:csd}), while the low-$T$ maximum indicates the 
            temperature of the structural
            transition from the isotropic to the orientationally ordered (anisotropic) system.
      }
      \label{torus:hc}
    \end{center}
  \end{figure}

\begin{figure}[t!]
    \begin{center}
	\centering
      \includegraphics[scale=1]{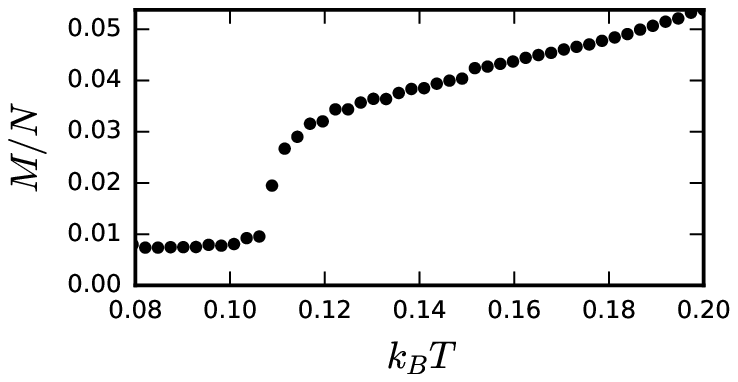}
            \includegraphics[scale=1]{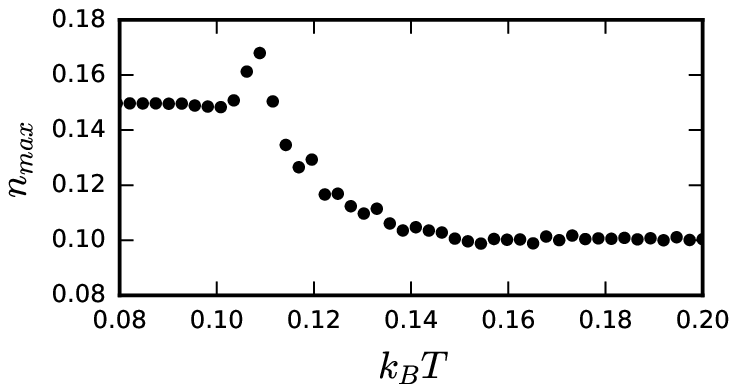}
            \caption{Properties of the aggregates as a function of temperature for the case of TBC. 
            {\color{black}{Left}} panel: the number of clusters, $M$, divided by the number of particles $N$. 
            {\color{black}{Right}} panel: the size of the largest cluster, $n_{max}$ divided by $N$. 
      }
      \label{torus:clusters}
    \end{center}
  \end{figure}

\begin{figure}[t!]
    \begin{center}
	\centering
      \includegraphics[scale=1]{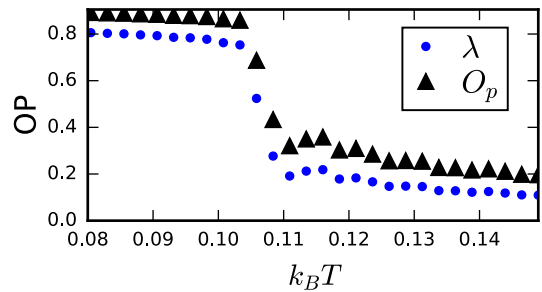}
      \caption{Orientational order parameters for the system with TBC.
      $\lambda$ is 
      the eigenvalue of the ordering matrix defined in Eq. (\ref{order_mat}),
      while  $O_p$ is the order parameter based on the moments of inertia ratio introduced in sec.\ref{sec:methods}.
      }
      \label{torus:op}
    \end{center}
  \end{figure}

Presence of TBC can lead to different structures in the SALR model depending on the system size. 
If the size is large enough, then the SALR 
particles form disordered isotropic stripe patterns at $k_BT > 0$, \cite{zarragoicoechea:09:0}
whereas in the case of smaller
systems the use of TBC can induce order \cite{imperio:06:0,almarza:14:0}. 
Typical configurations obtained in our simulations are 
presented in Fig. \ref{torus:snaps}. The high temperature structures show networks of particles that are structureless 
and disordered. The decrease of temperature results in formation of clusters with a preferred size, 
indicated by a local 
maximum in the ASD. Upon further cooling the preferred size increases slightly,
the clusters merge into longer structures 
and the ASD becomes bimodal (Fig. \ref{torus:csd}) due to formation of elongated aggregates, which we call stripes. 
Cluster merging is associated with a local maximum in the heat capacity (Fig. \ref{torus:hc}) and a monotonic
decrease of the number of the aggregates $M$  (Fig. \ref{torus:clusters}a). Interestingly, at temperatures $k_BT >0.14$, 
although the number of aggregates decreases, the size of the largest observed aggregate $n_{max}$ 
(Fig. \ref{torus:clusters}b) 
and the orientational order parameters (Fig. \ref{torus:op}) do not show any significant changes. 
Thus, we conclude that at the 
high temperatures the cluster formation and their merging into lamellar stripes leads to formation 
of isotropic structures 
made of easily distinguishable aggregates with a preferred and limited size. The size distribution of the aggregates 
(Fig. \ref{torus:csd}), however, is fairly broad, and thus the size fluctuations are large.

  \begin{figure}[t!]
    \begin{center}
\includegraphics[scale=1]{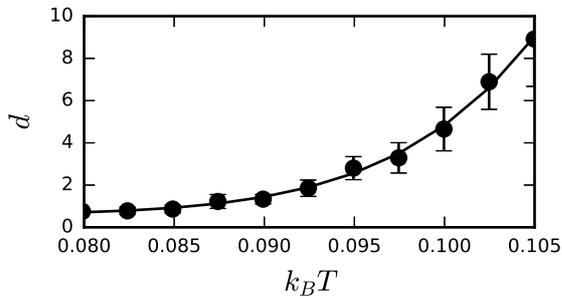}
            \caption{The number of defects as a function of temperature for the TBC case.
            The solid line is the fit of our results to the exponentially growing function of the form 
            $c+ a\exp(b/k_BT)$.
The number of dislocation defects is defined as a number of stripes ends divided by the factor of 2. 
The ends of the stripes were localized with the method described in Sec. \ref{sec:methods}.
      }
      \label{torus:defects}
    \end{center}
  \end{figure}
  
Further decrease of temperature triggers more intensive structural changes, which result in 
aligning of the lamellar 
stripes into orientationally ordered structure (Fig. \ref{torus:snaps}e-g). As the temperature decreases, the number 
of aggregates decreases and exhibits a significant drop at $k_BT \approx  0.11$. The rapid merging of clusters into stripes 
induces high energy fluctuations and as a result, another peak in the heat capacity occurs. In order to show that 
the structural transition that takes place at $k_BT \approx  0.11$ resembles an order-disorder phase transition, we calculated
two orientational order parameters described in Sec. \ref{sec:methods}. The first order parameter, $\lambda$, 
is the eigenvalue of the ordering matrix defined in Eq. (\ref{order_mat})  (Fig. \ref{torus:op}a) and the second one is the order parameter $O_p$  (Fig. \ref{torus:op}b). {\color{black} In fact, the two OP for the case of a flat two-dimensional systems are strictly related: $O_p = 1 - \frac{1-|\lambda|}{1+|\lambda|}$ }. However, only $O_p$ can be conveniently used for non-flat systems, as will be shown in Sec. \ref{sec:sphere}. 
Both order parameters exhibit a rapid change at $k_BT \approx 0.11$.  We verified that the transition occurred for 
different system sizes too, and that the temperature at the transition was the higher the smaller
was the system. We did not 
try to verify if this orientational ordering was associated with a thermodynamic phase transition
in the thermodynamic limit,
since in this work we are interested in comparing systems of finite area that are either flat, or curved and closed.

In the case of TBC, ordering of winding lamellar stripes into a defect-less lamellar structure
takes place in a narrow temperature range and is associated with merging of lamellar segments 
that are short. The emerging structure is composed of 7 stripes that thanks to periodicity of
the simulation box do not have topological defects and form closed curves. The structure, however,
is not translationally invariant probably due to incommensurability between the box size and the 
period of the structure or a mismatch between the box size and the number of particles.

Large slopes of the order parameters are  reflected in the birth rate of topological defects in the stripes.
The number of the defects 
in the structure is an important factor from the point of view of possible industrial applications. 
In the case of the diblock copolymers which spontaneously form lamellar structures, the number of dislocation defects
increases with temperature exponentially \cite{mishra:12:0} {\it via} the following relation 
\begin{equation}
n_d \sim \frac{1}{a_c^2} \exp^{-E_d /k_BT},
\label{expgrowth}
\end{equation}
where $n_d$ is the density of dislocations, $E_d$ is the energy of a single dislocation and $a_c$
is a dislocation core radius. 
In the case of the SALR particles, heating up the orientationally ordered
lamellar structure also results in an exponential growth
of the  number of the dislocation defects (Fig. \ref{torus:defects}).

\subsection{Spherical boundary conditions}
\label{sec:sphere}

 In this section, we study the system with the SBC and the same area as in the system with the
 TBC presented in Sec. \ref{sec:torus}. The density, however, had to be lowered because while the
 TBC allows the  SALR particles to form straight stripes, the SBC does not,
 and the packing of the stripes is less dense. 
 For this reason  in order to obtain ordered lamellar structures at low-T, a system with smaller density was used.

\begin{figure}[t]
    \begin{center}
	\centering
      \includegraphics[scale=1]{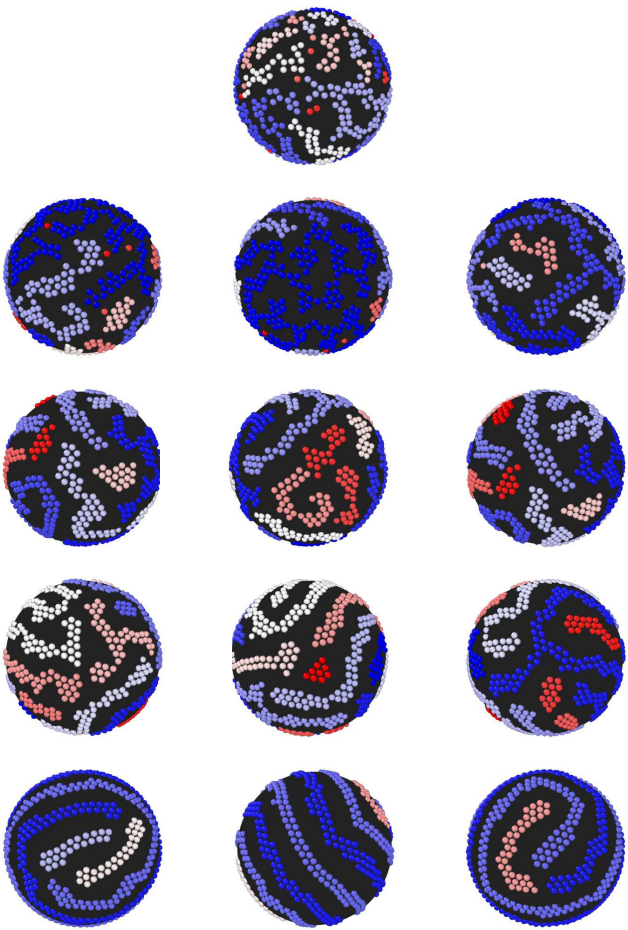}\\
            \caption{Representative configurations of the system with SBC and $N = 900$ at different temperatures.
            From top to bottom row:
            $k_BT = 0.18, 0.14, 0.11, 0.09, 0.06$. The number of clusters in the consecutive rows are:
            44, 24, 20, 17, 5. In each row different snapshots of the same configuration are presented.
            Different colors correspond to different clusters indentified by the distance criterium.
      }
      \label{sphere:snaps}
    \end{center}
  \end{figure}

\begin{figure}
    \begin{center}
	\centering
      \includegraphics[scale=1]{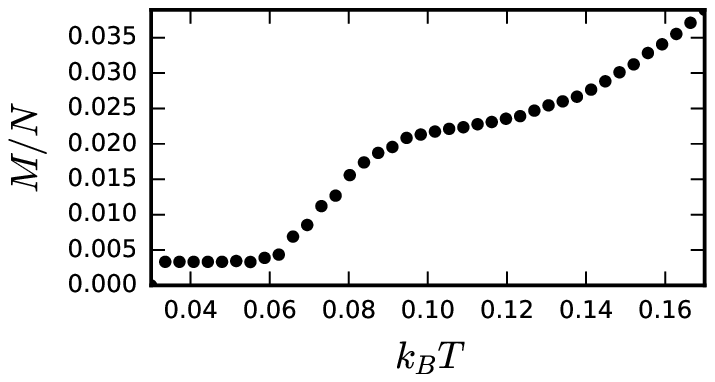}
      \includegraphics[scale=1]{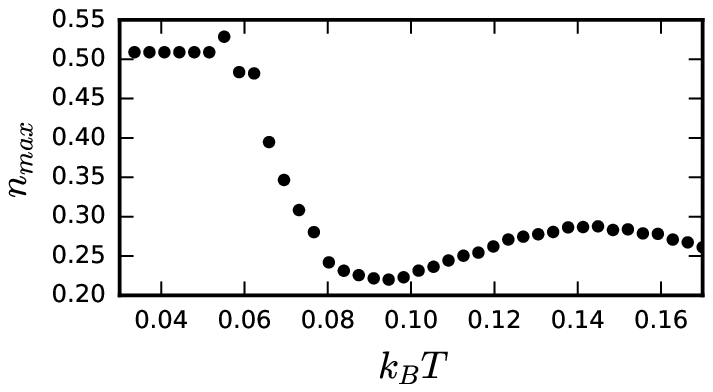}
      \caption{Cluster properties as a function of temperature for the case of SBC.
      The number of clusters ({\color{black}{left}} panel) and the number of particles in the largest cluster ({\color{black}{right}} panel) divided by the number of particles, $N$.
      }
      \label{sphere:clusters}
    \end{center}
  \end{figure}

\begin{figure}
    \begin{center}
\includegraphics[scale=1]{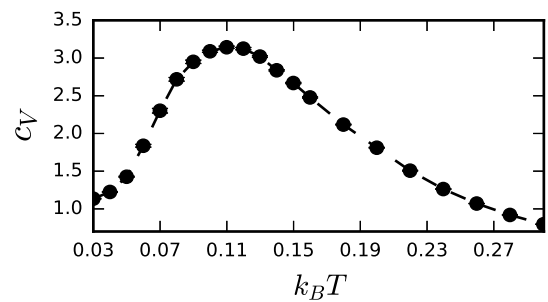}
            \caption{Heat capacity per particle for the system with SBC. The maximum is at $k_BT  = 0.11$.
      }
      \label{sphere:hc}
    \end{center}
  \end{figure}

\begin{figure}
    \begin{center}
	\centering
	\includegraphics[scale=1]{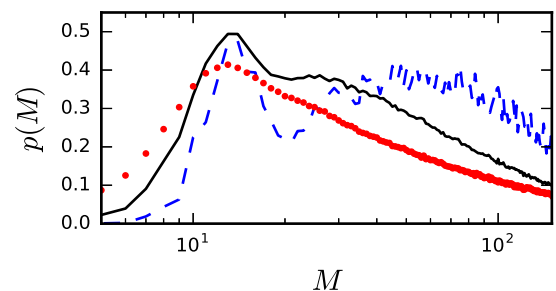}\\
            \caption{Mass weighted aggregate size distribution (Eq.(\ref{p(M)})) 
            for the spherical system with $N = 900$ particles
            at different temperatures. $k_BT = 0.08$ (dashed blue), $k_BT = 0.11$ (solid black),
            $k_BT = 0.14$ (dotted red).
      }
      \label{sphere:csd}
    \end{center}
  \end{figure}

 Representative configurations for different temperatures are shown in Fig. \ref{sphere:snaps}.
High temperature structural behavior of the system with the SBC resembles that of the system with the TBC.
The SALR fluid 
is disordered, but it is not homogeneous. When the temperature decreases, clusters with a preferred size start
to form. Similarly as in the case of the TBC, further cooling results in merging of the clusters 
(Fig. \ref{sphere:clusters}a) into elongated structures, but in the case of the SBC the $n_{max}$ 
is not constant and exhibits a local maximum at $k_BT \approx 0.14$ (Fig. \ref{sphere:clusters}b).
The maximum of $n_{max}$ occurs at the temperature at which also the slope of $M(T)$ changes and becomes smaller
for $k_BT<0.14$. When $k_BT$ is further decreased,
the clusters merge into stripes, and $M(T)$ exhibits an inflection point at  $k_BT \approx  0.11$. At 
the same temperature the heat capacity has a maximum (Fig. \ref{sphere:hc}) and the ASD becomes bimodal 
(Fig. \ref{sphere:csd}).

\begin{figure}[t!]
    \begin{center}
	\centering
      \includegraphics[scale=1]{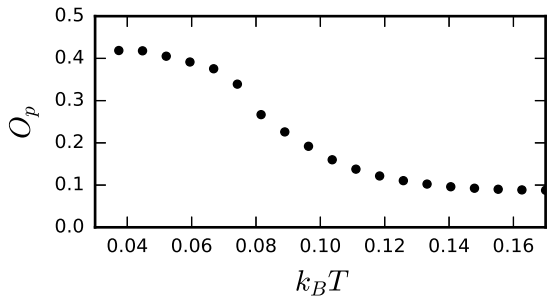}\\
                  \caption{Orientational order parameter $O_p$  defined in sec.\ref{sec:methods} for the system with SBC.
      }
      \label{sphere:op}
    \end{center}
  \end{figure}

\begin{figure}[t]
    \begin{center}
	\centering
      \includegraphics[scale=0.8]{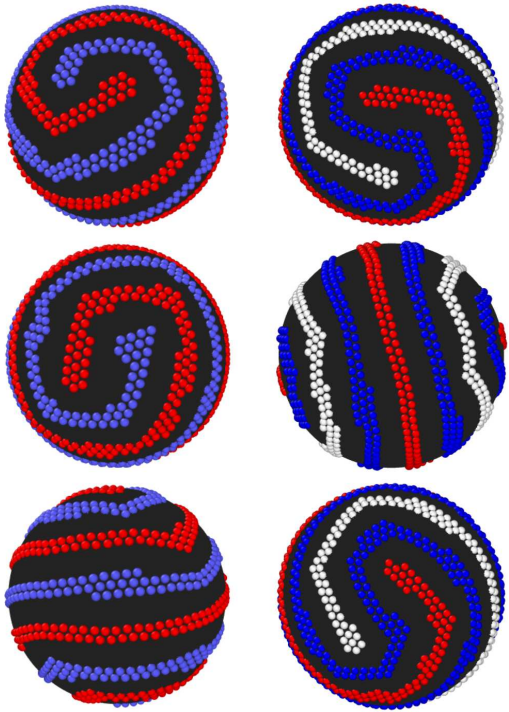}
            \caption{Low-temperature equilibrium structures. Left column:
            two lamellar stripes with two 
            topological defects, i.e. four ends of the stripes.  Right column:
             three stripes;   
           one of them forms a closed curve (marked by the blue color) and the other two (white and red) have four
            ends.
      }
      \label{sphere:gs}
    \end{center}
  \end{figure}

The orientational order parameter $O_p$ does not change as strongly as in the case of TBC (Fig.~\ref{sphere:op}).
The $O_p$ changes smoothly in the temperature range $0.06 < k_BT < 0.14$  between
 $O_p=0.09$ and  $O_p=0.42$. The low-T particle configurations are presented in Fig.~\ref{sphere:gs}.
The obtained structures are composed of either 2 or 3 lamellar stripes, which wrap around each other. 
In particular, the 2-stripe structure resembles a double-helix, while the 3-stripe structure is 
a double-helix with an extra stripe that separates the other two. Importantly, in both structures
only a local orientational order can be seen and both structures have similar total potential energy. 
Similar energies probably result from the same number of topological defects.

\begin{figure}
    \begin{center}
	\centering
\includegraphics[scale=1]{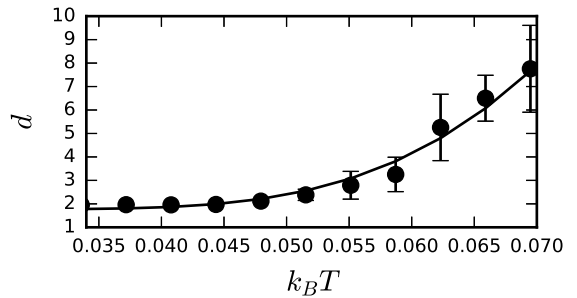}
                  \caption{The number of dislocations defects of the lamellar stripes for the system 
                  with SBC as a function of temperature. 
                   }
      \label{sphere:defects}
    \end{center}
  \end{figure}

Our results show that for relatively small SALR systems with spherical topology,
the number of dislocations increases with temperature {\it via} the exponential relation,
as it was in the case of TBC  (Fig. \ref{sphere:defects}).

Merging of lamellar segments into longer stripes upon cooling takes place in a much broader temperature range  then in the TBC case. Both $M/N$ and $n_{\text max}$ reach a plateau only for $k_BT \le 0.06$.

%Tracking how the average number of aggregates and their size changes with temperature 
%(Figs. \ref{sphere:clusters} and \ref{torus:clusters}) reveals significant differences between the TBC and SBC cases. 
%In the case of TBC, ordering of winding lamellar stripes into a defect-less lamellar structure
%takes place in a narrower temperature range and is associated with merging of lamellar segments 
%that are shorter. The emerging structure is composed of 7 stripes that thanks to periodicity of
%the simulation box do not have topological defects and form closed curves. The structure, however,
%is not translationally invariant probably due to incommensurability between the box size and the 
%period of the structure or a mismatch between the box size and the number of particles.

The heat capacity dependence on temperature is significantly different than in the case of the flat surface too.
In the case of the TBC there are two peaks: the high-T peak is associated with formation of the lamellar stripes,
while the low-T peak reflects the energy fluctuations rise due to the stripe ordering into defect-less structure. 
In the case of the SBC, only the stripe formation leads to an increase of the heat capacity, probably because 
the orientational ordering upon cooling occurs in a broad temperature range, the order 
is weak and topological defects are allowed.

\section{Discussion and conclusions}
\label{sec:summary}

 The aim of our study was to determine how the curvature of the surface at which self-assembling
 particles are adsorbed influences pattern formation. We have chosen the radius of the sphere
 one order of magnitude larger than the diameter of the adsorbed particles. The number of particles
 was chosen such that lamellar structures
 can be formed on both flat and curved surface.
 We have limited our study to a single value of the density because
 formation of lamellar structures at low $T$ requires very precise choice of the density~\cite{almarza:14:0}.
 
   To determine the effects of curvature,
 we have computed the number of aggregates $M(T)$, the maximal number of particels in the aggregate $n_{max}(T)$,
 the aggregate size distribution, the number of defects (half of the terminal points of elongated aggregates) $d(T)$, 
 the orientational
 OP $O_p(T)$,   and the specific heat $c_V(T)$, 
 %for a range of temperature
 for a flat and a spherical  surface, both of the same area.  
 
 For relatively high temperature the curvature does not play a significant role. The particles on 
 both, the flat surface and the surface of the sphere self-assemble into clusters when temperature is decreased.
 At some $T$ the clusters merge into elongated assemblies (stripes),
 and the aggregate size distribution becomes bimodal. The specific heat takes a maximum at this temperature.
The value of this temperature, however, is lower at the spherical surface.
 
 Further cooling of the system leads to merging of the short stripes into larger ones that tend to be
 parallel to each other. This process begins at similar temperatures in both cases. On the flat surface the 
 number of stripes decreases rapidly and the orientational OP increases 
 rapidly between $O_p\approx 0.3$ and $O_p\approx 0.85$ in the same very narrow temperature interval 
 around $k_BT\approx 0.11$. The 
 orientational ordering and the rapid decrease of the number of defects are accompanied by a pick
 in the specific heat. On the sphere, the increase of the $O_p$ and the decrease of the $M(T)$ occur
 much more gradually. Both parameters change in the  temperature interval $0.06<k_BT<0.12$. 
 In fact the upper boundary of the crossover region between the oriented and the isotropic structures
 cannot be uniquely
 determined, since the slope of $O_p(T)$ changes gradually for $0.06<k_BT<0.14$.
 The $O_p$ increases from $O_p\approx 0.1$ at $k_BT\approx 0.12$ to $O_p\approx 0.4$ at
 $k_BT\approx 0.06$ 
 that is much smaller than $k_BT\approx 0.11$ corresponding to the ordered structure on a flat surface of the same area. 
 Moreover, the broad crossover between the isotropic and anisotropic structures is not accompanied by
 a pick in the specific heat. Still, a significant difference between the high-$T$ isotropic structure with many defects
 and the low-$T$
 anisotropic structure with few defects can be seen.  
 We conclude that the curvature does not play a crucial role as long as the self-assembled 
 stripes are shorter than the radius of the sphere. When the length of the stripes becomes comparable with the 
 radius of the sphere or larger, then the curvature starts to play a significant role, as one should expect.

 In the anisotropic structure with few defects the number of defects  on the flat surface 
 increases according to eqn (\ref{expgrowth}) for increasing $T$, 
 and on the sphere the same relation, but shifted by the ground state value of 2, holds. 
 This is because according to our simulations the minimal number of defects on the sphere surface is $d=2$. 
 
 Let us discuss the low temperature structure on the sphere. One might naively expect that parallel
 rings
 %with the centers located at the same symmetry axis of the sphere and
 consisting of bilayers of particles would be formed. However, the lengths of the  two layers of particles in the ring
 are  different, 
 and this difference increases when the rings approach the poles of the sphere. 
 For this reason the distance between the particles in the two layers forming the ring
 are different, and it is not possible that most of the particles are separated by
 the distance corresponding to the minimum of the interaction potential.  
 Thus, the splay of the ring-forming stripe is associated with an energy cost which for our model is large.
In our model, the spirals that are parallel near the equator and make turns between segments that lie at a portion
of a big circle, and eventually near the pole of the sphere break (Fig.\ref{sphere:gs}), are energetically favourable.

It is interesting to compare the low-$T$ structures shown in Fig.\ref{sphere:gs} with the defects that occur in 
a nematic phase of rods adsorbed at a sphere. There are two  defects  
associated with the ends of the two open stripes. 
They  lie on the opposite sides of the big circle (or at the poles of the sphere),
in full analogy with the defects in the nematic phase.
(see  Fig.24 in Ref.\cite{bowick:09:0}).
Thus, the analogy between the ML phase in the SALR system and the nematic phase in a system of rod-like particles,
based on the orientational order in the two phases,
persists on the curved surface of the sphere.

\section{Acknowledgements}
This project has received funding from the European Union \textquotesingle s Horizon 2020 research and innovation programme under the Marie
Sk\l{}odowska-Curie grant agreement No 734276 (CONIN). An additional support in the years 2017-2018  has been granted  for the CONIN project by the Polish Ministry of Science and Higher Education. Financial support from the National Science Center under grant No. 2015/19/B/ST3/03122 is also acknowledged.

\end{document}